\documentclass[multphys,vecphys]{svmult}

\usepackage{makeidx}     
\usepackage{graphicx}    
\usepackage{multicol}    

\makeindex             

\begin{document}

\title*{The Nearby QSO Host I~Zw~1: NIR Probing of Structural Properties and
Stellar Populations}

\titlerunning{NIR Probing of the Nearby QSO Host I~Zw~1} 
\author{Julia Scharw\"achter\inst{1}\and
Andreas Eckart\inst{1}\and
Susanne Pfalzner\inst{1}\and
Jihane Moultaka\inst{1}\and
Christian Straubmeier\inst{1}\and
Johannes Staguhn\inst{2}\and
Eva Schinnerer\inst{3}}
\authorrunning{J. Scharw\"achter, A. Eckart, S. Pfalzner et al.}
\institute{I. Physikalisches Institut, Universit\"at zu K\"oln, Z\"ulpicher Str. 77, 50937 K\"oln, Germany
\texttt{scharw@ph1.uni-koeln.de}
\and NASA/Goddard Space Flight Center, Greenbelt, MD 20771, USA\texttt{}
\and National Radio Astronomy Observatory, PO Box 0, Socorro, NM 87801, USA\texttt{}
}
%
%
\maketitle

The likely merger process and the properties of the 
stellar populations in the I~Zw~1 host galaxy are analyzed on the basis of
multi-wavelength observations\footnote{With the 
ISAAC camera at 
the Very Large Telescope (VLT/UT1) of the European Southern Observatory 
(ESO), Chile (Paranal), 
with the interferometer of the Berkeley-Illinois-Maryland 
Association (BIMA), USA (Hat Creek/California), 
and with the IRAM 
Plateau de Bure Interferometer (PdBI), France\label{foot:1}} 
and N-body simulations. The data give a
consistent picture of I~Zw~1, with properties 
between those of ultra-luminous infrared
galaxies (ULIRGs) and QSOs as displayed by transition objects in
the evolutionary sequence of active galaxies.

\section{Introduction}
\label{sec:1}
The formation and evolution of active galactic nuclei (AGN) and the 
associated star-formation activity are conditioned by the rate at which 
gas is supplied from the surrounding host galaxy. Interactions and mergers are
identified as one possible mechanism to provide the transport of gas towards 
the central regions of a galaxy (\cite{1992ARA&A..30..705B} and references
therein). The hypothesis of a merger-driven evolution of active galaxies 
from ULIRGs to unobscured QSOs is motivated by the spectral energy
distributions of template objects showing a sequence of 
decreasing far-infrared excess and increasing UV brightness 
\cite{1988ApJ...325...74S}. 
In accordance with the assumption that the evolution is triggered by 
interactions, the vast majority of ULIRGs show evidence for ongoing
merger processes \cite{1999Ap&SS.266..331S}. 
Merger signs are less clear in the case of objects close to the QSO phase. 
These transition objects are important sources 
to confirm the evolutionary sequence. 
As a nearby representative ($z=0.0611$, \cite{1997ApJ...478..144S}) of 
a template transition object in the original evolutionary sequence 
(Fig.~17 in \cite{1988ApJ...325...74S}), I~Zw~1
is suitable to address the merger properties, the structural 
composition, and the starburst activity 
using near-infrared
(NIR) and molecular line observations at sub-kpc angular resolution.

\section{Observations and Results}
\label{sec:2}

\subsection{Merger Process}
\label{sec:2.1}
\begin{figure}
\centering
\includegraphics[width=10cm]{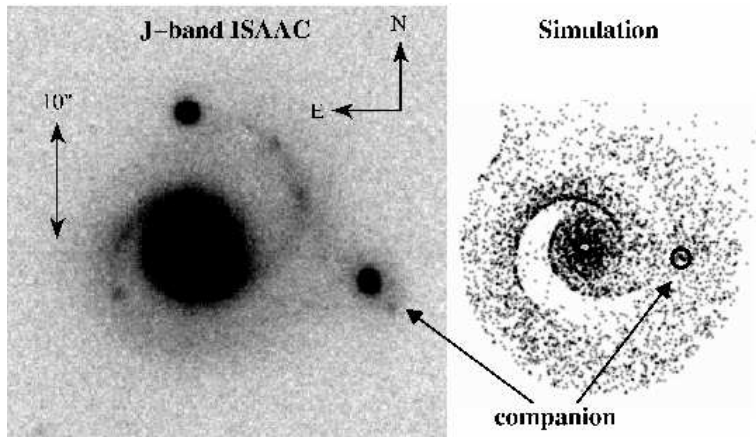}
\caption{J-band image of I~Zw~1 (left panel), taken with the ISAAC camera at 
the VLT of ESO, and a snapshot of a restricted N-body simulation for 
the interaction between I~Zw~1 and the companion (right panel). The companion
in the simulation has a mass of 3\% of the total mass of I~Zw~1 and orbits 
I~Zw~1 at an eccentricity of 0.2 and with a pericentric distance of 15.3~kpc.}
\label{fig:1}       
\end{figure}

The J-band image of I~Zw~1 (Fig.~\ref{fig:1}), 
obtained with ISAAC at a median seeing of 0.6'' 
($\approx 0.7$~kpc), provides a detailed view
of the host galaxy with its two-armed
spiral, its likely western companion galaxy, and the northern foreground star.
The ongoing merger process with the companion is indicated by 
the tidal bridge towards the companion and by the tidal tail and 
the elongated shape of the companion \cite{2003A&A...405..959S}. 
The parameter space of this interaction is investigated with the help of 
restricted stellar-dynamical N-body 
simulations. As suggested by the surface-brightness profile of the companion,
the study focuses on companions with less than 10\%
of the total mass of I~Zw~1. On a bound elliptical orbit, even a companion 
with only 3\% of the total mass of I~Zw~1 can evoke a two-armed
spiral structure similar to the one observed in the J-band image
(Fig.~\ref{fig:1}).

\subsection{Hydrogen Emission and Stellar Absorption Lines}
\label{sec:2.2}
\begin{figure}
\centering
\includegraphics[width=10cm]{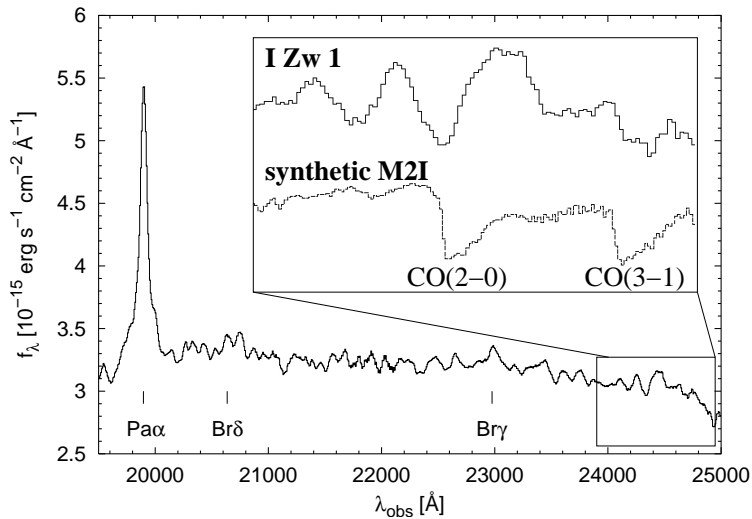}
\caption{K-band spectrum of I~Zw~1 taken with a 1'' long-slit and extracted
in a 3'' aperture centered on the I~Zw~1 nucleus. Stellar absorption 
features are magnified in the inset
and compared to the synthetic spectrum of an M2 super-giant.}
\label{fig:2}       
\end{figure}

The K-band spectrum of I~Zw~1 shown in Fig.~\ref{fig:2} was obtained with 
the 1'' long-slit of ISAAC and
extracted in a 3'' aperture centered on the I~Zw~1 nucleus. It is
characterized by strong 
hydrogen emission lines (Pa~$\alpha$, Br~$\gamma$, Br~$\delta$).
In agreement with I~Zw~1 being classified 
as a narrow-lined Seyfert~1, 
the hydrogen lines are narrow with a full width at half maximum of 
about $1,000$~km~s$^{-1}$. The narrow lines with broad wings
indicate a mixture of emission from narrow-line regions and 
from broad-line
regions in the surroundings of the AGN. The flux ratio of the Pa~$\alpha$ and
the Br~$\gamma$ line is about 11, as expected for HII regions.
The overall fluxes and line widths are in good agreement with the results 
from previous K-band spectroscopy of I~Zw~1 \cite{1998ApJ...500..147S}. 
There are indications for the $^{12}$CO(2-0) 
bandhead (inset in Fig.\ref{fig:2}) which is 
mainly coming from late-type giants and/or super-giants.
The preliminary analysis of the depth of this bandhead suggests a stellar
contribution of about 20\% to the total nuclear NIR flux of I~Zw~1.
This is in good agreement with the results presented in 
\cite{1998ApJ...500..147S}, which are 
based on the analysis of the $^{12}$CO(6-3) overtone bandhead 
in the H-band spectrum of I~Zw~1. 
By comparing stellar and dynamical masses, it becomes evident that
such a high stellar flux contribution can only be produced by a massive
starburst (see \cite{1998ApJ...500..147S}).

\subsection{Circum-Nuclear Molecular Gas}
\label{sec:2.3}
The circum-nuclear $^{12}$CO(1-0) emission in I~Zw~1, 
observed with the BIMA
interferometer at an 
angular resolution of 0.7'' ($\approx 0.8$~kpc),  
has a ring-like structure with a radius of about 1.5~kpc 
(see \cite{2001IAUS..205..340S}). 
The comparison of the BIMA
$^{12}$CO(1-0) measurements with PdBI
$^{12}$CO(2-1) data indicates that the molecular gas
in the disk region is cold or subthermally excited while the gas in the 
circum-nuclear region is warm and optically thick. The large molecular clouds
in the ring-like structure are most probably the sites of the massive 
starburst activity discussed in \cite{1994ApJ...424..627E}, \cite{1998ApJ...500..147S}, and Sect.~\ref{sec:2.2}.

\subsection{Structural Decomposition and Mean Stellar Populations}
\label{sec:2.4}
Mass-to-light (M/L) ratios 
of the mean stellar populations in the bulge
and disk components of the I~Zw~1 host are derived by decomposing the 
J-band surface-brightness profile and the 
gas rotation curve (see \cite{2003A&A...405..959S}, 
\cite{2003Ap&SS.284..507S}, 
\cite{2001qhte.conf..289S}). 
The disk component and, in particular, 
the bulge component, show sub-solar M/L ratios. 
In accordance with the findings of starburst activity, 
the low M/L ratios indicate young mean stellar populations
with an enhanced fraction of 
hot stars and super-giants.

\section{Conclusion}
\label{sec:3}
The multi-wavelength data converge to a consistent picture of 
I~Zw~1 as a transition object. A merger-driven
evolution is suggested by the J-band image and the N-body simulations 
showing evidence for a merger process with the companion. 
The decomposition, 
together with the large stellar contribution to the nuclear NIR flux,
indicates young mean stellar populations and starburst activity
which is most likely located in the giant clouds of the circum-nuclear molecular ring.

\printindex
\end{document}